\documentclass
[preprint,a4paper,byrevtex,nofootinbib,nobibnotes,noshowpacs,aps]{revtex4}%
\usepackage{amsfonts}
\usepackage{amssymb}
\usepackage{amsmath}
\usepackage{graphicx}%
\begin{document}
\author{Xin Zhou*\footnotetext{* Corresponding author. E-mail: xinzhou@wipm.ac.cn;
Tel: +86-27-8719-7796; Fax: +86-27-8719-9291.}, Jun Luo, Xian-ping Sun, Xi-zhi
Zeng, Ming-sheng Zhan, Shang-wu Ding, Mai-li Liu}
\affiliation{State Key Laboratory of Magnetic Resonance and Atomic and Molecular Physics,
Wuhan Institute of Physics and Mathematics, The Chinese Academy of Sciences,
P. O. Box 71010, Wuhan 430071, People's Republic of China}
\title{Experiment and Dynamic Simulations of Radiation Damping of Laser-polarized
liquid $^{129}$Xe at low magnetic field in a flow system}

\begin{abstract}
Radiation damping is generally observed when the sample with high spin
concentration and high gyro-magnetic ratio is placed in a high magnetic field.
However, we firstly observed liquid state $^{129}$Xe radiation damping using
laser-enhanced nuclear polarization at low magnetic field in a flow system in
which the polarization enhancement factor for the liquid state $^{129}$Xe was
estimated to be 5000, and furthermore theoretically simulated the envelopes of
the $^{129}$Xe FID and spectral lineshape in the presence of both relaxation
and radiation damping with different pulse flip angles and ratios of
T$_{2}^{\ast}$/T$_{rd}$. The radiation damping time constant T$_{rd}$ of 5 ms
was derived based on the simulations. The reasons of depolarization and the
further possible improvements were also discussed.\newline

\end{abstract}
\maketitle

\newpage

\section{Introduction}

Laser-polarized $^{129}$Xe and $^{3}$He gases \cite{happer} have been found
wide applications in polarized targets \cite{xu}, magnetic resonance imaging
(MRI) \cite{albert,ebert}, neutron polarization \cite{Jones}, fundamental
symmetry studies \cite{chupp}, high resolution nuclear magnetic resonace (NMR)
spectroscopy \cite{raftery}, surface science \cite{pietrass}, precision
measurements \cite{bear}, biological-system probe \cite{rubin}, quantum
computer \cite{Verhulst}\ and cross polarization to other nuclei
\cite{driehuys,long}. The enhanced NMR signals of laser-polarized $^{129}$Xe,
which are about 10$^{5}$ times larger than those from thermally polarized
$^{129}$Xe \cite{happer,Sun}, have opened the possibility to explore radiation
damping effect \cite{1} of laser-polarized $^{129}$Xe.

Since Bloembergen and Pound analyzed the physical process of radiation damping
in 1954 \cite{1}, radiation damping had been noticed and studied in the early
time \cite{2,3,4,5}. Although the research of radiation damping in NMR had
almost halted from 1960 to 1988, it has once again evoked the interest of
researchers in recent years because of the applications of high field magnets
and high sensitive probes. Since then, a large number of research articles and
reviews have been published \cite{6,7,8,9,10,11,12,13,14,15}.

Radiation damping is a non-linear effect. The current, induced by the
transverse magnetization in the receiver rf coil, interacts with the
magnetization vector itself, tending to bring it back to the +Z axis (the
direction of the static magnetic field). Radiation damping is usually observed
at high fields where liquid phase magnetization \textit{M}, the static
magnetic field homogeneity, and the quality factor \textit{Q} of the probe are
high enough \cite{1,2,6}. The magnetization is \textit{M=}$\gamma\hslash
$\textit{CP} in the light of the Currie law, with $\gamma$ the gyromagnetic
ratio, \textit{P} the polarization, and \textit{C} the spin concentration. In
aqueous solutions the water proton concentration is usually 110 mol/L
(\textit{C}$\approx$110 M), which is high enough to produce radiation damping
effect. Radiation damping can also be observed for other solvents and
concentrated samples \cite{ref}. According to the thermal-polarization
\textit{P}$_{0}$\textit{(B}$_{0}$\textit{,T)}$\approx\gamma$\textit{hB}$_{0}%
$\textit{/(4}$\pi$\textit{kT)}, either decreasing the temperature or
increasing the magnetic field can enhance the polarization and make radiation
damping effect stronger, but it can provide only limited relief. It's very
natural to employ optical pumping and spin exchange \cite{happer} to improve
the polarization.

Although radiation damping of laser-polarized gaseous xenon was obtained at 14
T on Bruker DRX 600 spectrometer \cite{CPL-99}, the proper inversion of the
magnetization could not be entirely intepreted with the dipolar field theory
\cite{Warren_Sci}. The dipolar field effects \cite{Jeener_PRL_82}, on the
other hand, could explain NMR instabilities and spectral clustering in
laser-polarized liquid xenon \cite{PRB_V63}. In medical MRI, the existence of
radiation damping both in the phantom experiments and in vivo affects the
quality of imaging \cite{ZhouJinyuan}, and one can forsee that the same
problem will emerge in application of the lung MRI using hyperpolarized $^{3}%
$He and $^{129}$Xe \cite{albert}. Therefore, until now, the microcosmic
mechanism of radiation damping have be still not fully understood, and the
corresponding dynamics problems need to be resolved.

Two recent articles \cite{19,20} have reported radiation damping of
laser-polarized $^{3}$He, which is easier to be observed than that of $^{129}%
$Xe because the gyromagnetic ratio $\gamma$ and optical pumping efficiency of
$^{3}$He is larger than that of $^{129}$Xe. Here we report first observation
of radiation damping in laser-polarized liquid $^{129}$Xe at the low magnetic
field in a flow system, and the typical radiation damped FID envelopes and
spectra at different pulse flip angles are presented and compared with
theoretical ones. Characterization of such a system is important for
hyperpolarized MRI at low magnetic fields \cite{JMR_157_235}. Our experiment
could provide another liquid phase example to anlysis radiation damping
effect. In addition, laser-polarized liquid xenon and its radiation damping
also have a number of potential applications, such as construction of liquid
xenon maser \cite{Rosenberry}, rapid and precise measurement of liquid xenon
polariztion \cite{JMR_155}, etc..

\section{Experiment}

The diagram of our experimental apparatus is shown in Figure 1. The pump cell
containing a few drops of metal Cs was maintained at approximately 333$\pm1$ K
by the resistance heater during optical pumping. The inner surfaces of the
cylindrical Pyrex tube and the pump cell were coated with silane in order to
slow down the relaxation of the $^{129}$Xe upon collision with the tube wall.
The cell was placed in a 25 Guass magnetic field generated by Helmholtz coils.
The whole system was evacuated with K2 valve close and K1, K3 and K4 valves
open. When the vacuum reached 1.5$\times$10$^{-5}$ Torr, the valves K1, K4
were closed and K2, K3 were opened. The cell was filled with 740 Torr natural
xenon gas (26\% enriched $^{129}$Xe gas). After all the valves were closed,
laser light from a 15 W laser array (Opto Power Co. Model OPC-D015-850-FCPS)
at 852.1 nm was introduced to the system. The laser light resonates with the
Cs D$_{2}$ transition line and induces an electron spin polarization in the Cs
vapor via a standard optical pumping process \cite{21}. The hyperpolarized
$^{129}$Xe gas was produced by spin-exchange collision at the same time. The
polarization process took about 25 minutes. K4 valve was then opened to allow
the $^{129}$Xe to be transmitted into the probe, pre-cooled to 172$\pm1$ K, of
Bruker SY-80M NMR spectrometer. The temperature of the probe was,
subsequently, reduced to 160$\pm1$ K to freeze the xenon into condensed state.
But the actual temperature, which was somewhat different from the monitor's,
was improved from NMR observation that solid phase and liquid phase xenon
co-exist in the probe because the temperature rang between the melting point
and the boiling point of xenon was small (4.6$\pm3$ $^{0}$C). To study
radiation damping of the liquid $^{129}$Xe, the sample was cooled down to
142$\pm1$ K to completely freeze the xenon, and then gradually warmed up and
maintained at 166$\pm1$ K.

\section{Results and Discussion}

Figure 2 shows the $^{129}$Xe NMR time domain signal (a) and the corresponding
spectrum (b) at the temperature of 160$\pm1$ K when 160$^{0}$ pulse was
applied. The peak with larger line width at high frequency ($\delta$147) is
assigned to solid $^{129}$Xe and the sharp peak with phase distortion
($\delta$108) is from liquid $^{129}$Xe (Fig. 2b). Such phase distortion is
the typical result of the radiation damping.

It should be noted that at the temperature of 166$\pm1$ K, only liquid state
$^{129}$Xe NMR signal at $\delta$108 is observable. Figure 3 shows the
$^{129}$Xe FID signals (a1-a4) and the corresponding frequency spectra (b1-b4)
excited by 60$^{0}$, 90$^{0}$, 120$^{0}$, and 150$^{0}$ pulses, respectively.
Compared with the water theoretical FID envelope and spectrum in Figure 4
\cite{3,9}, these signals obviously indicated that the laser-polarized liquid
$^{129}$Xe have radiation damping effects. Unlike conventional FID (Fig. 3
a1-a2), the profile of typical radiation damped FID excited by a pulse flip
angle greater than 90$^{0}$ will increase initially, and start to decay after
reaching a maximum, and the position of the maximum is flip angle dependent
(Fig. 3 a3-a4). As it is expected, the amplitude of symmetrical phase twisting
of the corresponding spectra is enhanced when the flip angle is increased
(Fig. 3 b3-b4) \cite{10,11}.

The strength of radiation damping is characterized by radiation damping time
constant \textit{T}$_{rd}$ \cite{2}. The longitudinal relaxation time
(\textit{T}$_{1}$) of $^{129}$Xe in liquid phase is about 30 minutes
\cite{22,23}. The \textit{T}$_{2}^{\ast}$ of $^{129}$Xe in the NMR machine
(Bruker SY-80M ) due to the poor inhomogeneity is estimated to be 30 ms.
Radiation damping can occur only when \textit{T}$_{rd}$ is shorter than
\textit{T}$_{2}^{\ast}$ \cite{10}. Compared with \textit{T}$_{2}^{\ast}$ and
\textit{T}$_{rd}$, \textit{T}$_{1}$ effect is negligible. The value of
\textit{T}$_{rd}$ can be estimated from the line-width of spectrum obtained by
using very small pulse flip angle, since in this case the line shape will be
reduced to a Lorentzian \cite{11}.
\begin{equation}
\Delta v_{1/2}=q/\pi T_{2}^{\ast}=\pi^{-1}[(1/T_{2}^{\ast})+(1/T_{rd})]
\end{equation}
Considering the error of $\Delta v_{1/2}$ and \textit{T}$_{2}^{\ast}$, the
values of \textit{T}$_{rd}$ was in the range of 6$\pm$2ms. Then, according to
the analytical results of the Bloch equations including only \textit{T}$_{rd}$
and \textit{T}$_{2}^{\ast}$, which have been presented in the Mao's paper
\cite{11},
\begin{equation}
S(\omega)=(2M_{0}T_{rd}q/T_{2}^{\ast})\sum(-1)^{n}\frac{(2n+1)q/T_{2}^{\ast}%
}{[(2n+1)q/T_{2}^{\ast}]^{2}+\omega^{2}}[\tan(\eta/2)]^{2n+1},(for:0<\eta
\leq\pi/2);
\end{equation}%
\begin{equation}
(2M_{0}T_{rd}q/T_{2}^{\ast})\sum(-1)^{n}\frac{(2n+1)q/T_{2}^{\ast}%
}{[(2n+1)q/T_{2}^{\ast}]^{2}+\omega^{2}}\{2\cos(\omega t_{0})-[\cot
(\eta/2)]^{2n+1}\},(for:\pi/2<\eta\leq\pi).
\end{equation}
with
\begin{equation}
q=[1+(T_{2}^{\ast}/T_{rd})^{2}+2(T_{2}^{\ast}/T_{rd})\cos\theta_{0}]^{1/2},
\end{equation}%
\begin{equation}
t_{0}=-(T_{2}^{\ast}/q)\tanh^{-1}\{[(T_{2}^{\ast}/T_{rd})\cos\theta
_{0}+1]/q\},
\end{equation}%
\begin{equation}
\cos\eta=[(T_{2}^{\ast}/T_{rd})\cos\theta_{0}+1]/q,
\end{equation}
we simulated the spectra with different \textit{T}$_{rd}$ values and pulse
flip angles in order to get the accurate \textit{T}$_{rd}$ value. And the same
simulation of the FID was performed using the formula \cite{11}:
\begin{equation}
M_{y}=(M_{0}T_{rd}/T_{2}^{\ast})q\sec h[(q/T_{2}^{\ast})(t-t_{0})].
\end{equation}
\textit{T}$_{2}^{\ast}$ of liquid state $^{129}$Xe was assumed to be 30 ms in
all simulations unless otherwise indicated, which was the same as the
experimental one. As a result, we found the precise \textit{T}$_{rd}$ of
laser-polarized liquid $^{129}$Xe was about 5ms in our experiment. The
simulated $^{129}$Xe FID and corresponding frequency spectrum, assumed the
150$^{0}$ pulse and \textit{T}$_{rd}$=5ms, are respectively visualized in
Figure 5 (a) and (b), which are in good agreement with our experimental
results [Fig. 3 (a4) and (b4)].

Figure 6 shows the theoretically simulated envelopes of the $^{129}$Xe FID in
the presence of relaxation and radiation damping with different flip angle
pulses and \textit{T}$_{rd}$ values. From the spectra, we can learn that the
envelope of the FID is associated with the ratio of \textit{T}$_{2}^{\ast}%
$\textit{/T}$_{rd}$ and the flip angle $\theta_{0}$. If \textit{T}$_{2}^{\ast
}$\textit{/T}$_{rd}\leq$1, the bigger the ratio is, the faster the FID decays
when $\theta_{0}\leq$90$^{0}$, contrary to the cases of $\theta_{0}$%
$>$%
90$^{0}$; and the radiation damping impossibly occurs and the maximum
amplitude point is located at t=0 whatever the flip angle $\theta_{0}$ is. If
\textit{T}$_{2}^{\ast}$\textit{/T}$_{rd}$%
$>$%
1, the same situation occurs that the bigger the ratio is, the faster the FID
decays; but when $\theta_{0}$%
$>$%
90$^{0}$ the FID grows up at first, then falls down after reaching the maximum
amplitude point, which is higher when the ratio is bigger. The spectra also
indicate that the FID decays monotonically when $\theta_{0}\leq$90$^{0}$ no
matter what the ratio is. The radiation damping can be observed only when
\textit{T}$_{2}^{\ast}$\textit{/T}$_{rd}$%
$>$%
1 and when the pulse flip angle is greater than 90$^{0}$. The bigger
\textit{T}$_{2}^{\ast}$\textit{/T}$_{rd}$ is, the stronger radiation damping
will be. The radiation damping will also occur even when \textit{T}$_{2}%
^{\ast}$\textit{/T}$_{rd}$=2. So our observation of the liquid $^{129}$Xe
radiation damping under the conditions of the large $^{129}$Xe magnetization
and the ratio of \textit{T}$_{2}^{\ast}$(30ms)/\textit{T}$_{rd}$(5ms)=6 is reasonable.

Based upon the relation of the magnetization versus the nuclear spin
polarization given by Abragam \cite{5}, and by comparison of the nuclear spin
polarization \textit{P}$_{L}$ of $^{129}$Xe produced by laser optical pumping
with the Boltzmann polarization \textit{P}$_{B}$ of $^{129}$Xe at the thermal
equilibrium, the enhancement factor of the $^{129}$Xe nuclear spin
polarization can be expressed as:
\begin{equation}
f=\frac{P_{L}}{P_{B}}=\frac{I_{L}}{I_{B}}%
\end{equation}
where \textit{I}$_{L}$ and \textit{I}$_{B}$ are the integral intensities of
the measured NMR signals of the liquid $^{129}$Xe under the conditions of the
laser optical pumping and the thermal equilibrium, respectively. Therefore,
the enhancement factor of the liquid $^{129}$Xe was about 5000 on the basis of
our measurement \cite{Zhou_Acta}. The Boltzmann polarization \textit{P}$_{B}$
of the $^{129}$Xe at 166 K in the Bruker SY-80M (1.879 Tesla) spectrometer is
about 2.9$\times$10$^{-6}$, hence the liquid state $^{129}$Xe nuclear spin
polarization \textit{P}$_{L}$ is 1.45\%, which was produced by spin exchange
with laser-polarized Cs atoms at the low magnetic field of 25 Guass in a flow system.

From the qualitative face, radiation damping emerges when the transverse
relaxation time \textit{T}$_{2}^{\ast}$ is longer than the characteristic time
of radiation damping \textit{T}$_{rd}$=$\frac{1}{2\pi\eta QM\gamma}$
\cite{1,2,6}, although the quality factors \textit{Q} of our Bruker SY-80M NMR
spectrometer is rather poor and the filling factor is rather small, and the
radiation damping effect is about four times more efficient in the case of
protons than for $^{129}$Xe nuclei at the same magnetization level because of
the dependence of \textit{T}$_{rd}$ on the gyromagnetic ratio. But when the
$^{129}$Xe nuclear polarization is enhanced, the signal intensity of liquid
$^{129}$Xe is larger than that of water protons in a high magnetic field by
comparing water protons signals at 11.7 Telsa and room temperature
(\textit{C}=110 M, \textit{P}=4.2$\times$10$^{-5}$, \textit{M}=7.83$\times
$10$^{-5}$ G/cm$^{3}$) with laser-polarized liquid $^{129}$Xe signals
(\textit{C}=5.46 M, \textit{P}=1.45$\times$10$^{-2}$ , \textit{M}=3.69$\times
$10$^{-4}$ G/cm$^{3}$). The value of \textit{T}$_{2}^{\ast}$ will dramatically
be shortened due to the magnetic field inhomogeneity, so radiation damping of
water protons could not be observed on the same SY-80M spectrometer, but with
the enhancement of polarization, we have observed the liquid phase radiation
damping for our laser-polarized sample. The phenomenon, that radiation damping
occurs for liquid $^{129}$Xe but not for solid $^{129}$Xe under the same
polarization condition, also demonstrates that radiation damping depends
primarily on both the intensity of the magnetization and the characteristic
time of radiation damping \textit{T}$_{rd}$.

The polarization in a flow system is not so high as that in the close pump
cell or produced by the narrow bandwidth Ti:Sapphire laser \cite{CPL-99}. The
main causes are: (a) The loss of polarization is due to the magnetic field
inhomogeneity during the transfer of laser-polarized $^{129}$Xe gases. (b)
Since the bandwidth of our diode laser array is 30 $\overset{\circ}{A}$ but
the absorbed bandwidth of Cs atom in the 740 Torr natural xenon is only 30
GHz, the efficient pump power is only about 0.125 W. (c) The phase transition
of laser-polarized $^{129}$Xe can also bring about the depolarization. (d) The
relaxation of $^{129}$Xe atoms at the walls still decreases the polarization
even though they were coated.

Thus we can see the further possible improvements in our experiment by
adopting the following techniques: (a) Creating the homogeneous magnetic field
by two-layer solenoid with end correction coils, and keeping the transmission
of laser-polarized xenon paralleling the static magnetic field. (b) Increasing
the gas pressure of the pump cell in order to enhance the optical-pumped
absorbed power. (c) Decreasing the time of phase transition of laser-polarized
$^{129}$Xe.

\section{Conclusions}

Usually radiation damping was observed only when the sample has high spin
concentration, such as water protons, in the high magnetic field with
high-resolution spectrometer. However, our experiment indicated that under the
condition of laser-enhanced nuclear polarization, which made the liquid
$^{129}$Xe polarization enhancement of 5000 compared to that without optical
pumping under the same conditions, the liquid $^{129}$Xe radiation damping can
also be observed even at low magnetic field in a flow system. The liquid
$^{129}$Xe \textit{T}$_{rd}$ of our experiment was estimated to be as short as
5ms with the help of radiation damping line shape theory, and we theoretically
simulated the FID and spectrum of $^{129}$Xe in the presence of radiation
damping and the transverse relaxation in order to compare with the
experimental ones. We also discussed whether the radiation damping would occur
with different ratios of \textit{T}$_{2}^{\ast}$\textit{/T}$_{rd}$, and the
theoretically simulated envelopes of the $^{129}$Xe FID in the presence of
both relaxation and radiation damping with different pulse flip angles and
\textit{T}$_{rd}$ values were presented. The reasons of depolarization and the
further possible improvements were also discussed. Furthermore, our experiment
of laser-polarized liquid $^{129}$Xe NMR may be a parttically good test bed
for studying micro-mechanism of radiation damping.

\section{Acknowledgments}

This work is supported by the National Natural Science Foundation of China
under Grant No. 10234070, National Science Fund for Distinguished Young
Scholars under Grant No. 29915515 and National Fundamental Research Program
under Grant No. 2001CB309306. One of the author (X. Z.) is grateful to Dr.
Xi-an Mao for helpful discussions and suggestions.

\begin{center}
{\LARGE Figure captions}
\end{center}

Fig. 1 The diagram of our experimental apparatus.

Fig. 2 The experimental $^{129}$Xe FID signals that the liquid phase and the
solid phase coexisted with radiation damping when the 160$^{0}$ [(a)] pulse
was applied and the corresponding frequency spectrum [(b)].

Fig. 3 The experimental liquid phase $^{129}$Xe FID signals when 60$^{0}$
[(a1)], 90$^{0}$ [(a2)], 120$^{0}$ [(a3)] and 150$^{0}$ [(a4)] pulses were
applied and the corresponding frequency spectra [tagged with (b1,b2,b3,b4)], respectively.

Fig. 4 The envelope of FID with different flip angles under strong radiation
damping [(a)]. The theoretical lineshape of water strong radiation damping
with the 150$^{0}$ flip pulse excitation, in which the T$_{rd}$ of water
proton was assumed to be 12ms[(b)].

Fig. 5 The theoretically simulated $^{129}$Xe FID [(a)] and corresponding
frequency spectrum [(b)] when the 150$^{0}$ pulse was applied with T$_{2}^{*}%
$=30ms, and T$_{rd}$=5ms.

Fig. 6 The theoretical envelopes of the $^{129}$Xe FID in the presence of
relaxation (T$_{2}^{*}$=30ms) and radiation damping while the different pulse
flip angles and T$_{rd}$ were assumed: T$_{rd}$=60ms [(a)], T$_{rd}$=30ms
[(b)], T$_{rd}$ =15ms [(c)], and T$_{rd}$=5ms [(d)].
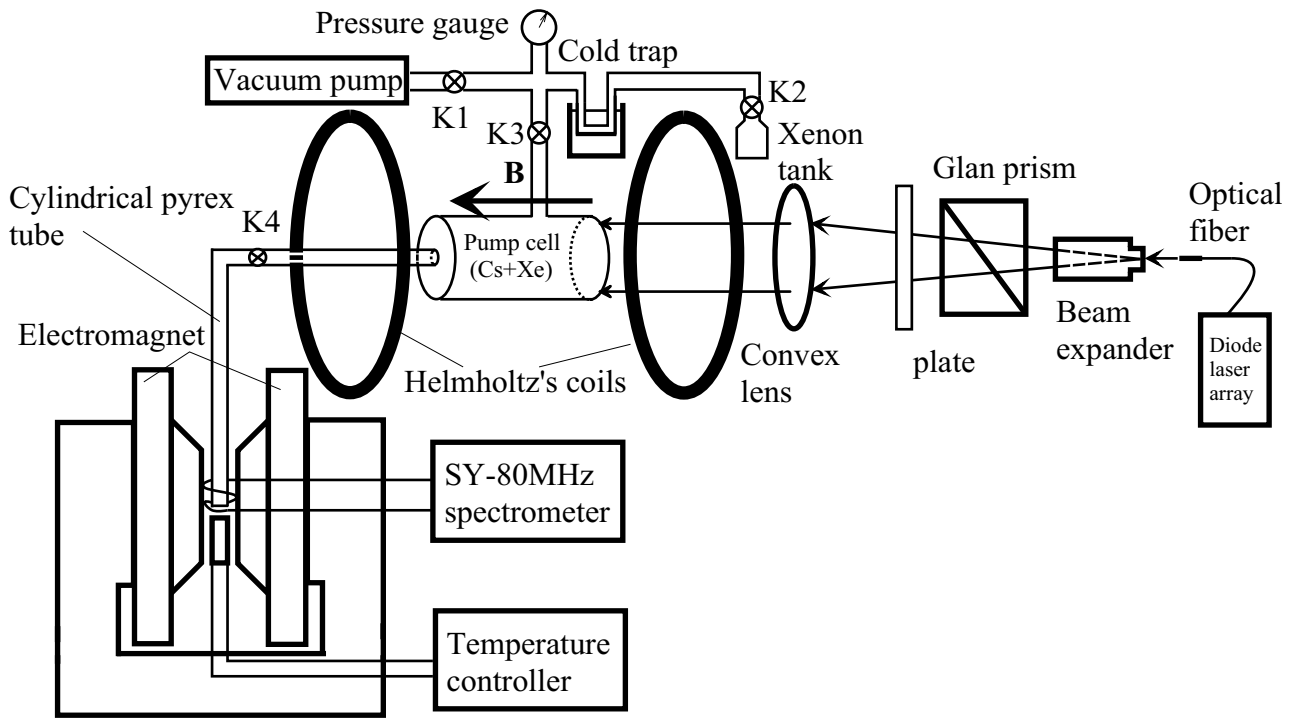
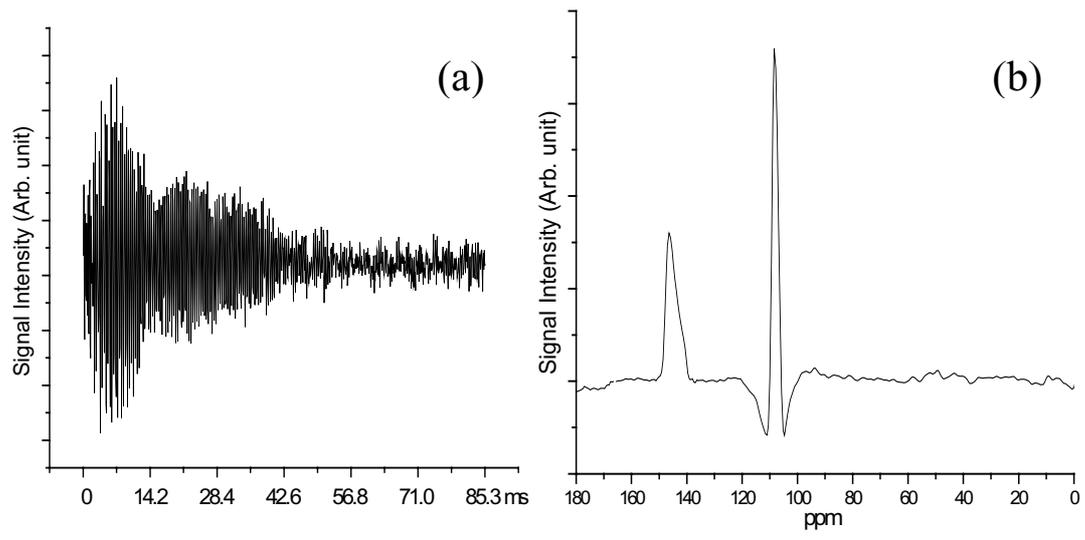
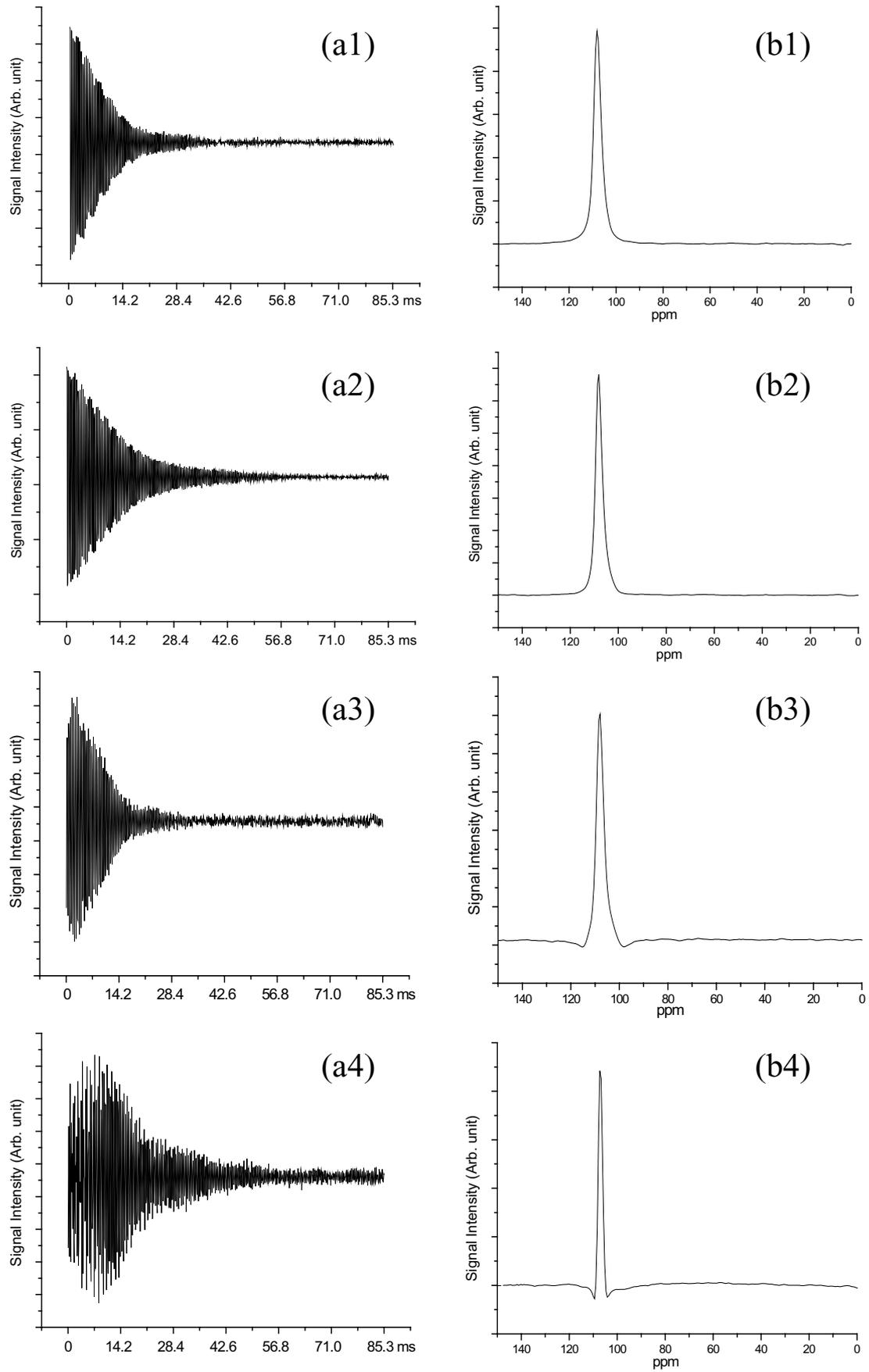
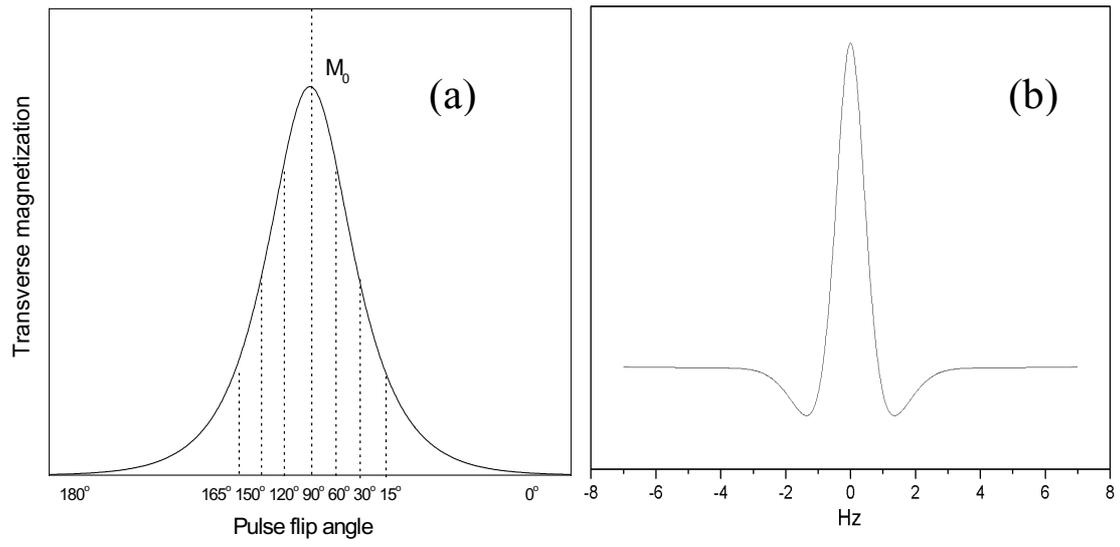
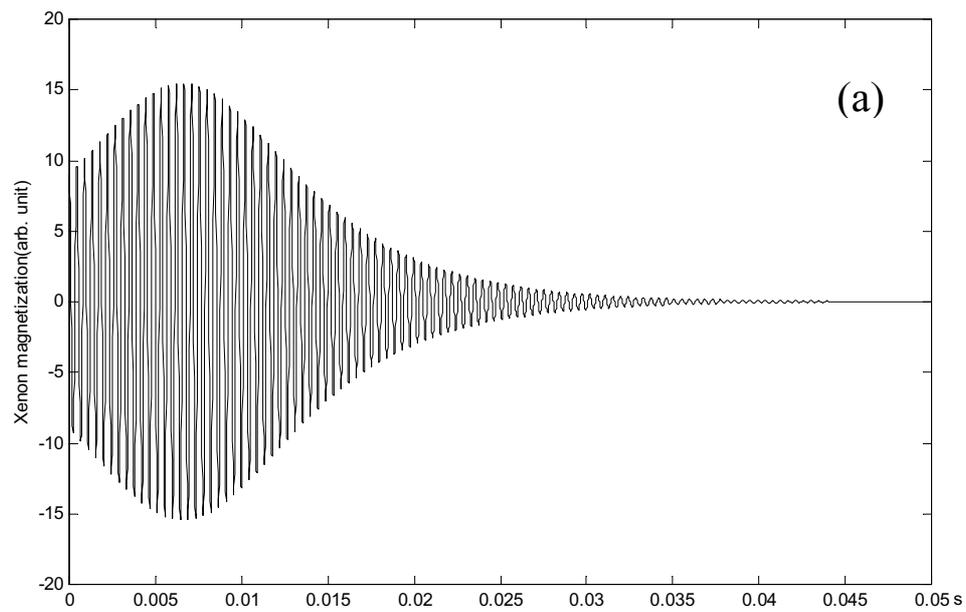
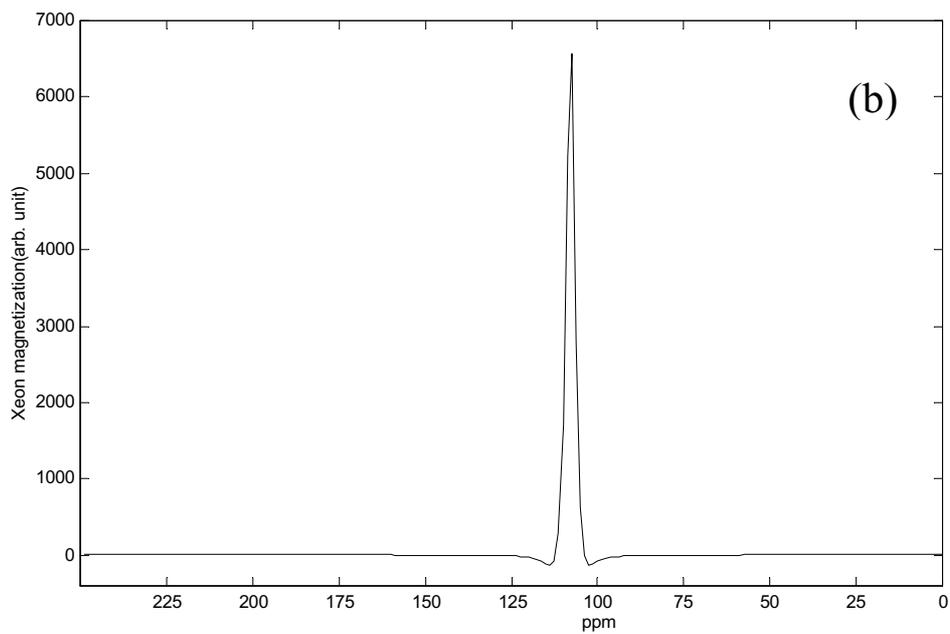
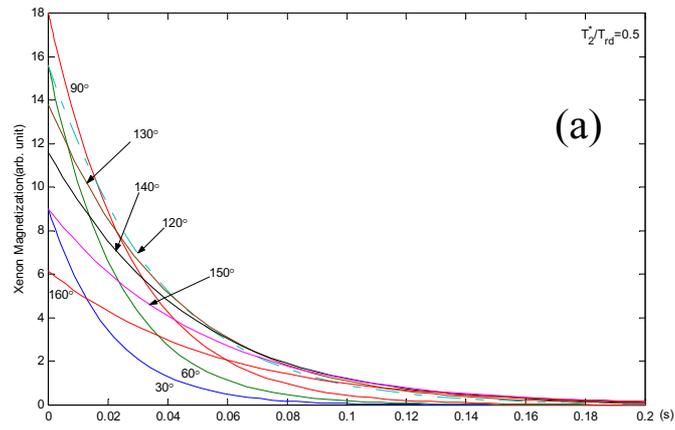
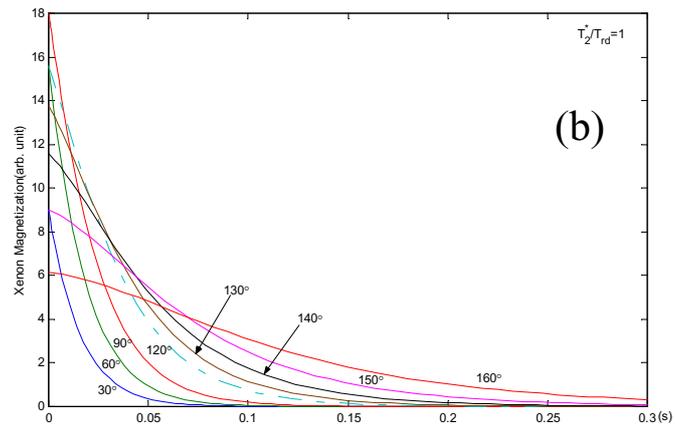
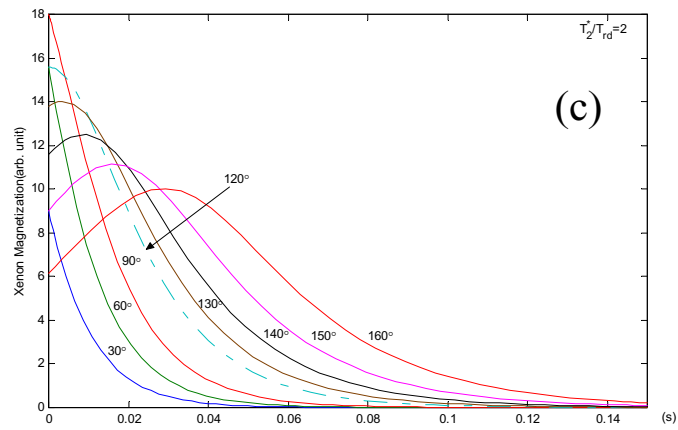
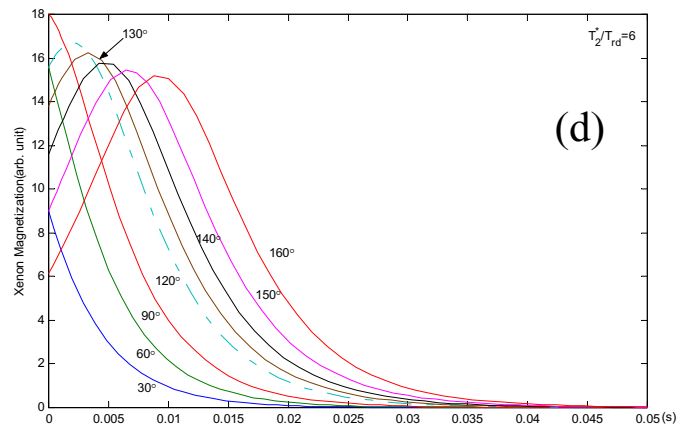
\end{document}